\documentclass[page-classic]{epl2}

\shorttitle{Comment on ``$^3$He mass diffusion"}
\shortauthor{Lamoreaux and Golub} \institute{
  \inst{1} Yale\\
  \inst{2} NCSU
} 
 \abstract{We
clarify a point regarding $^3$He--$^3$He collisions in our
recent measurement.}

\begin{document}

\title{Comment on ``Measurement of the $^3$He mass diffusion coefficient in
superfluid $^4$He over the 0.45---0.95 K temperature range"}
\author{S.K Lamoreaux\inst{1} \and R. Golub\inst{2}}
\maketitle

In the analysis of our measurement of the mass diffusion of
$^3$He in He II,\cite{1} we assumed that the heat flow and the
$^3$He concentrations were sufficiently small so that the liquid
could be assumed as isothermal. We also assumed that the
``relatively infrequent" $^3$He-$^3$He collisions did not
influence the steady-state $^3$He atom distribution imposed by
the heat flow. This latter assumption requires justification.
These collisions are in fact irrelevant to the problem.

In our experiment, we were studying the equilibrium steady-state
distribution of the $^{3}$He atoms. In such a case, the $^{3}$He
atoms are at rest on average (no diffusion or momentum current), while the
normal fluid (phonons), which carries heat, is moving past them.
The normal fluid
velocity $v_{n}$ exerts a force on a $^{3}$He atom that is given
by the mobility, $\vec{v}_{n}=\mu \vec{F}$. Thus the problem is
better treated through use of the mobility $\mu $ of a $^{3}$He
atom in a phonon field\cite{3}, which is related to the phonon
diffusion coefficient through the Einstein relationship,
\begin{equation}
D={k_{b}T\mu }.  \label{one}
\end{equation}
Because the mobility describes the effects of external forces on
a gaseous ensemble of particles, the response cannot be
influenced by self-collisions as these preserve the total
momentum of the gas. Therefore Eq. (\ref{one}) does not contain
any effects of $^{3}$He--$^{3}$He collisions. In other words
there is no $^3$He current so the $^{3}$He--$^{3}$He collisions
do produce a net force on the $^3$He. But the phonon wind
creates a force that can be derived from a potential because the
velocity field is irrotational.

Rewriting Eq. (5) of \cite{1}, the $^{3}$He atom spatial
distribution is described by
\begin{equation}
X=Ce^{\Phi /D}=Ce^{\Phi /\mu k_{b}T}
\end{equation}%
where $\vec{\nabla}\Phi =v_{n}$, the normal fluid component
velocity, and $C$ is a normalization constant. This equation is
simply the Boltzmann factor that describes the spatial
distribution of particles in a potential, at a fixed
temperature. Again, the fact that the $^{3}$He atoms are
colliding with each other does not alter the their Boltzmann
distribution in the potential, because momentum is conserved in
these collision. The implication is that $\Phi /\mu $ is a true
potential, and represents the work required to move a $^{3}$He
atom at one point in the measurement cell, to another, against
the normal velocity flow (or phonon current). For this to be
valid, the concentration $X$  needs to be low enough so that
potential due to $^{3}${He--}$^{3}$He interactions does not
cause an additional effective potential when the spatial $^3$He
distribution changes due to the heat flow.  This condition was
met in the experimental study,  $X\leq 10^{-4}$; such effects
might be expected at $X\geq 0.05$, e.g, where degenerate Fermi
gas effects come into play.


\end{document}